\documentstyle[prl,aps,epsf,multicol]{revtex}

\newcommand{\bleq}{\ifpreprintsty
                   \else
                   \end{multicols}\vspace*{-3.5ex}{\tiny
                   \noindent\begin{tabular}[t]{c|}
                   \parbox{0.493\hsize}{~} \\ \hline \end{tabular}}
                   \fi}
\newcommand{\eleq}{\ifpreprintsty
                   \else
                   {\tiny\hspace*{\fill}\begin{tabular}[t]{|c}\hline
                    \parbox{0.49\hsize}{~} \\
                    \end{tabular}}\vspace*{-2.5ex}\begin{multicols}{2}
                    \fi}
\newcommand{\bcols}{\ifpreprintsty\else\begin{multicols}{2}\fi}
\newcommand{\ecols}{\ifpreprintsty\else\end{multicols}\fi}

\begin{document}
\draft

\title{Multiple Attractor Bifurcations: A Source of Unpredictability in
Piecewise Smooth Systems}
\author{Mitrajit Dutta~\cite{MDandEO}
 \and Helena E. Nusse~\cite{HNandJY}~\cite{onlyHN}
 \and Edward Ott~\cite{MDandEO}~\cite{onlyEO}
 \and and James A. Yorke~\cite{HNandJY}~\cite{onlyJY}}
\date{\today}
\address{University of Maryland, College Park, Maryland 20742}
\maketitle
\begin{abstract}

There exists a variety of physically interesting situations
described by continuous maps that are nondifferentiable on some surface in
phase space. Such systems exhibit novel types
of bifurcations in which multiple coexisting attractors can be created
simultaneously. The striking feature of these bifurcations is that they lead
to fundamentally unpredictable behavior of orbits as a system parameter is
varied slowly through its bifurcation value. This
unpredictability gradually disappears as the speed of
variation of the system parameter through the bifurcation is
reduced to zero. 

\end{abstract}


\bcols

     In the literature dealing with bifurcation theory, differentiable maps
are commonly considered. On the other hand, maps that are piecewise
differentiable occur in a variety of physical situations. We will consider
here two dimensional maps. By a piecewise differentiable two dimensional map
we mean that the map is continuous and there is a curve (which we call the
{\em border}) separating the phase space into two regions, such that 
the map is differentiable on both sides of the border, but not on it.
In particular, in this letter we shall be concerned with the case where
the map's derivative changes discontinuously across the border. This
circumstance leads to a rich class of bifurcation phenomena 
called {\em border-collision bifurcations}~\cite{ny92,noy94}, examples of which
may be found in instances of grazing impact in mechanical 
oscillators~\cite{n91,fb92,bd94,lb94,cong94}, in piecewise-linear electronic
circuits~\cite{oi94}, and in commercial power electronic 
devices~\cite{l81,ck96}.
 
    In this letter, we study the occurrence and prevalence of border-collision 
bifurcations in which multiple coexisting attractors are simultaneously
created at the
bifurcation.  We call these bifurcations {\em multiple attractor
bifurcations}.

    Our principal point in this letter is that multiple attractor bifurcations
display
a striking new type of extreme sensitivity to noise. In particular, assume that
the noise $\delta N$ is at all times limited by some small value $\epsilon >
|\delta N|$. Assume
that before the bifurcation, the orbit follows a particular attractor (to
within the noise)
and then imagine that, as the system orbit evolves, the bifurcation parameter
is very slowly varied, eventually passing through its value at the bifurcation. 
The question is: After the bifurcation, which attractor
does the orbit follow? We argue that no matter how small the noise level
$\epsilon$ may be, this question may be fundamentally unanswerable.

	As a simple representative example, we consider the electronic circuit
with an analog switch~\cite{switch}, $S$, 
controlled by a comparator~\cite{compar}, $O$, as depicted in
Fig. 1a. Let $q$ denote the electric charge on the capacitor $C$, and let $i$
denote the current flowing through the inductor $L$ and the resistor $R$.
For small values of the current, $i$, the voltage drop, $iR$, across the
resistor $R$ is smaller
than the reference voltage, $v_{ref}$. Consequently, the comparator, $O$, keeps
the switch, $S$, open (off). However, when the current increases so that
$i > i_{crit} = v_{ref}/R$, the comparator output changes state, thereby closing
the switch, $S$. Ideally, the transition of the switch from an open state to a 
closed state (and vice versa), should be sharp and instantaneous. In reality,
however, the situation is somewhat compromised by a variety of factors including
finite gain and slew rate in the comparator op-amp, and finite rise and fall 
times of the switch. This causes a slight {\em blurring} of the $i = i_{crit}$
boundary. Fortunately, the precision and speed afforded by present-day 
components
restricts this boundary broadening to levels insignificant compared to the 
typical value of $i$ in the circuit, thereby justifying the sharp boundary 
model. The entire system is driven by an externally applied voltage
signal $v(t)$ which is a square wave of amplitude $v_{amp}$ with a constant 
bias $v_{bias}$. The circuit model can be described by the following normalized equations:
\begin{eqnarray}
        d Q/d T & = & \left\{ \begin{array}{lcr}
                                  I & & (I \leq 1)\\
                                  I-(Q+Q_{bias})/\rho & & (I > 1)
                                              \end{array} \right. \\
        d I/d T & = & \left\{ \begin{array}{lcr}
                   -\Omega^2Q -\Gamma I +F & & (0<T \leq 1/2)\\
                   -\Omega^2Q -\Gamma I -F & & (1/2<T \leq 1)
                                              \end{array} \right.
\end{eqnarray}
where $I = i/i_{crit}$, $Q = (fq/i_{crit}) - Q_{bias}$, and the time $T$ is
normalized to the period of the square wave drive, $T = ft$ where $f$ is the
drive frequency. The parameters in (1) and (2) are $\rho = fCR_C$, $F = v_{amp}
(fLi_{crit})^{-1}$, $\Omega^{-1} = f\sqrt{LC}$, $\Gamma = R/(fL)$, and $Q_{bias}
= v_{bias}fC/i_{crit}$.
Figures 2(a) and (b) show bifurcation diagrams for this system for the
two-dimensional time-1 map of this system (for parameters see the figure
caption). The interesting fact is that when
(as was done to get Figs. 2)
one creates bifurcation diagrams
by following the attractor 
as the bifurcation parameter is varied, upon addition of even
extremely small noise
(the level of which can be judged from the small thickening of the lines in the 
bifurcation diagram), we find that after the bifurcation the orbit can go to
different attractors: either a period three attractor (Fig. 2(a)) or a chaotic
attractor (Fig. 2(b)). 

	Before proceeding further, we introduce a canonical form~\cite{y97}
for border-collision bifurcations given in~\cite{ny92}, that allows us to treat
piecewise-differentiable systems in a general {\em system-independent} manner.
Let $G_{\mu}$  be a one-parameter family of piecewise smooth maps from the
two dimensional phase space to itself, depending smoothly on the parameter 
$\mu$. Let $E_{\mu}$ denote a fixed point (period one orbit) of $G_{\mu}$
defined on $\mu_0 -\alpha < \mu < \mu_0 +\alpha$ and depending continuously on
$\mu$, for some $\alpha > 0$ (The analysis may be easily extended to include
points on a period-$p$ orbit by choosing the map to be the $p$-th iterate of
$G_{\mu}$). As the parameter $\mu$ is increased from $\mu_0 -\alpha$, we suppose
that the fixed point $E_{\mu}$ collides with the border at $\mu = \mu_0$. 
We could also allow the border to depend on 
$\mu$, but coordinates can always be chosen so that it remains fixed.
For $\mu$ near $\mu_0$ the attractors are located near $E_{\mu_0}$, and one can
expand the map to first order for $(x, y)$ near $E_{\mu_0}$ and $(\mu-\mu_0)$
small. Nusse and Yorke~\cite{ny92}
show that after suitable changes of coordinates and rescaling of the
parameter, the resulting 
continuous, piecewise linear map can be cast in the form
\bleq
\begin{equation}
\hat{G_\mu}(x_n, y_n) = \left( \begin{array}{c} x_{n+1} \\ y_{n+1} \end{array}
\right) = \left[ \begin{array}{rr} \tau(x) & 1 \\ -d(x) & 0 \end{array} \right]
\left( \begin{array}{c} x_n \\ y_n \end{array} \right) + \left( \begin{array}{c}
\mu \\ 0 \end{array} \right),
\end{equation}
\eleq 
\noindent
where $\tau(x)$ and $d(x)$ are piecewise constant with a discontinuity at
$x = 0$, that is, $(\tau(x), d(x)) = (\tau_<, d_<)$ for $x \leq 0$ and
$(\tau(x), d(x)) = (\tau_>, d_>)$ for $x > 0$ where $\tau_<, d_<, \tau_>$, and
$d_>$ are constants. In Eq. (3) the border has been transformed
to the $y$-axis, $\mu_0 = 0$ and $E_{\mu_0} = (0, 0)$. We refer to (3) as the
{\em canonical form} of the original map $G_{\mu}$. By the piecewise
linearity of (3) in $(x, y)$ and its linearity in $\mu$, it is invariant under
$(x, y, \mu) \rightarrow (\lambda x, \lambda y, \lambda \mu)$. Thus,
as $\mu$ is reduced towards zero, any attractor of the system 
and its basin of attraction must contract
linearly with $\mu$, with the attractor
collapsing to the point $(x, y) = (0, 0)$ (cf., Figs. 2(c)
and (d) in
which both the period three attractor and the chaotic attractor collapse onto
the point $(0, 0)$). Thus it appears that, 
no matter how small the noise level $\epsilon$
may be, there always exists a finite, positive $\mu$-interval where the 
attractors are close enough for the noise to induce inter-attractor hopping,
thereby rendering prediction of the final attractor 
impossible~\cite{ott-fundae,extra}.

The canonical form (3) depends on four parameters 
$\tau_<, d_<, \tau_>$, and $d_>$ (and the bifurcation parameter $\mu$). To find
them from the original map $G_\mu$, we evaluate its Jacobian
matrix of partial derivatives for $\mu = \mu_0$ on both sides of the border,
infinitesimally close to $E_{\mu_0}$.
The parameters $\tau_<, \tau_>$ and $d_<, d_>$ are then the trace and
the determinant of the Jacobian matrix on the two sides of the border. As an
example, we evaluate the canonical form for the system, Eqs. (1, 2),
with parameters
corresponding to those of Figs. 2(a) and 2(b). Figures 2(c) and 2(d) show
bifurcation diagrams obtained from this canonical form (map $\hat{G_\mu}$) 
along with some very small additive noise.
Note that this figure reproduces the particular bifurcation phenomenon: there is
a border-collision bifurcation of the fixed point attractor ($\mu < 0$) to two
attractors ($\mu > 0$), one a period three attractor and the other a chaotic
attractor. This is an example of a multiple attractor bifurcation.

     In our example, Eqs. (1, 2), the Jacobian determinant is constant,
$d_< = d_>
= d > 0$, and for simplicity we henceforth limit our considerations to this
case. Thus, for fixed $d$, the type of border-collision bifurcations that occurs
depends on the two parameters $\tau_<$ and $\tau_>$. We wish to explore this 
parameter space to assess the prevalence of multiple attractor bifurcations.
Figure 3 shows the ($\tau_<, \tau_>$)-plane for $0.5 < \tau_< < 0.9$,
$-2.2 < \tau_> < -1.8$, $d = 0.3$ and $\mu = 1$.
Two specific initial conditions, $(x_0, y_0) = (7.0, 2.0)$ and 
$(x_0, y_0) = (1.2, 0.0)$, were chosen for our simulations at every 
($\tau_<, \tau_>$) grid point. The results show four distinct regions --- two
where both the initial conditions lead to a period-3 attractor, another
where both the initial conditions lead to a chaotic attractor, and still
another where one of the initial conditions leads to a period-3 attractor while
the other leads to a chaotic attractor. Clearly, the fourth region represents
parameter values where at least two attractors co-exist after bifurcation, thus
making multiple attractor bifurcations possible.

	It has been shown so far how a novel form of unpredictability
might arise in piecewise smooth maps. However, this new uncertainty gradually
gives way to certainty as the bifurcation speed (the rate at which the
bifurcation parameter is varied in time) is reduced to arbitrarily low
values. In order to understand this
consider the case where there are two attractors $A$ and $B$, and $\mu$
and the upper bound $\epsilon$ on the noise level are held fixed. For very
small $\mu$, the (noiseless) attractors are closer together than the noise level
$\epsilon$. For sufficiently large $\mu$ the attractors are so far apart
that the noise cannot kick an orbit on one of the attractors out of its basin,
and the orbit stays near the same attractor, $A$ or $B$, forever. In general,
however, there will be intermediate $\mu$ values $\mu_A$ and $\mu_B$ such that
hopping from $A$ to $B$ becomes impossible when $\mu > \mu_A$, and
hopping from $B$ to $A$ becomes impossible when $\mu > \mu_B$. Generically
(in the absence of special symmetries) $\mu_A \neq \mu_B$, and we assume 
$\mu_B > \mu_A$; e.g., attractor $B$ might be closer to the basin boundary than 
is attractor $A$.
Therefore, if the bifurcation parameter were to be varied infinitely
slowly, during the time when $\mu_B > \mu > \mu_A$, the system would certainly
end up in $A$ and it would remain there.
Thus, in the quasistatic
limit, there is no unpredictability as to which attractor the system ends
up in.

	Thus unpredictability occurs at nonzero bifurcation speed and we now ask
whether we can predict anything about the
relative probabilities of ending up on the different attractors given
the bifurcation speed.

	For simplicity, let us consider the situation with just
two post-bifurcation attractors, $A$ and $B$, as described in a previous
paragraph. Let $v$ denote the bifurcation speed, while $a(\mu)$ and $b(\mu)$
represent
the probabilities of the system being in attractors $A$ and $B$ respectively.
Then for small $v$ we have
\begin{eqnarray}
a + b & = & 1,\\
d b/d t & = & -\lambda_{BA} b + \lambda_{AB} a,
\end{eqnarray}
where $\lambda_{AB}(\mu)$ and $\lambda_{BA}(\mu)$ are the noise induced 
transition probabilities from $A$ to $B$ and $B$ to $A$, respectively.
Since $\lambda_{AB}(\mu) = 0$ for $\mu > \mu_A$, we immediately have
\begin{equation}
b(\mu) = b(\mu_A) e^{-\frac{1}{v}\int_{\mu_A}^{\mu_B}{\lambda_{BA}(\mu')d \mu'}},
\end{equation}
for $\mu > \mu_B$, where we assume that we begin at some $\mu < \mu_A$.
It should be pointed out that as $v \rightarrow 0$, $b(\mu_A) \rightarrow 0$ as 
well, but only as $v^\beta$ for
$\lambda_{AB}(\mu \rightarrow \mu_A^-) \sim (\mu_A - \mu)^\beta$. 
Therefore, to
leading order, $b(\mu > \mu_B)$, the probability of the system ending up on the
non-quasistatic attractor, satisfies $\ln [b(\mu > \mu_B)] \sim -1/v$.

	Two sets of simulations were conducted for Eq. (3) at the
parameter values corresponding to the electronic circuit described 
before. The system was subjected to additive noise with bounded
amplitude. In one case, the noise vector distribution had the form of a
uniformly filled square centered at the origin. In the other case, the
distribution was a uniformly filled circular disk also centered at the origin.
In both cases, 
when the speed of variation of the bifurcation parameter was sufficiently small,
the system tended to relax to the period three attractor.
Figure 4 confirms that, at intermediate speeds, the probability of not going to
the period three attractor indeed conforms to Eq. (6).

     In conclusion, in this paper, we have discussed a novel form of uncertainty
that can arise in piecewise smooth systems. We have also pointed out how
that uncertainty slowly fades into certainty as the speed of bifurcation
is reduced to zero.

     The computer assisted pictures were made by using 
{\em Dynamics}~\cite{ny97}. This work was in part supported by the Department of
Energy, the W.M. Keck Foundation, the ONR and the NSF.

\ecols

\pagebreak
\section*{Figure Captions}
\subsection*{Figure 1}
\noindent (a) The Border Collision Circuit.
\newline \noindent (b) $i \leq i_{crit}$.
\newline \noindent (c) $i > i_{crit}$.
\subsection*{Figure 2}
\noindent (a) and (b) Bifurcation diagrams for the two-dimensional time-1 map of
the circuit
depicted in Fig. 1 
with $Q_{bias} = -1.0$, $\rho = 0.10742$, $\Omega = 1.0642$ and $\Gamma = 
1.2040$. In creating the bifurcation diagrams (a) and (b) small noise was
inserted
leading to two different realizations.

\noindent (c) and (d) Bifurcation diagrams for the equivalent canonical form,
Eq. (3), near criticality
with $\tau_< = 0.7$, $\tau_> = -2.0$ and $ d = 0.3$. In creating (c) and (d),
small noise was inserted leading to two different realizations.
\subsection*{Figure 3}
\noindent Two-dimensional period plot in parameter space for the canonical form,
Eq. (3),
with $0.5 < \tau_< < 0.9$, $-2.2 < \tau_> < -1.8$, $d = 0.3$, $\mu = 1$, and
initial conditions $(x_0, y_0) = (7.0, 2.0)$ and $(x_0, y_0) = (1.2, 0.0)$.

\subsection*{Figure 4}
\noindent Probability $b(\mu>\mu_2)$, plotted logarithmically,
as a function of $1/v$, based on 1,000,000 experiments for every value of $1/v$.
Solid lines and plus symbols represent {\em square} noise, while dashed lines
and asterisks represent {\em circular} noise.

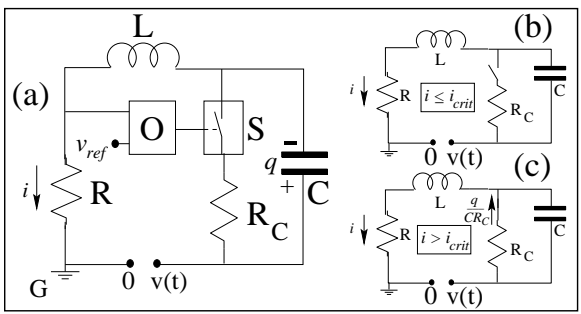
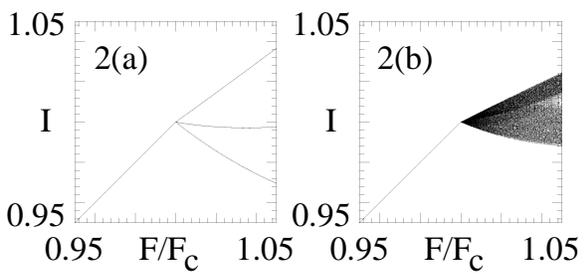
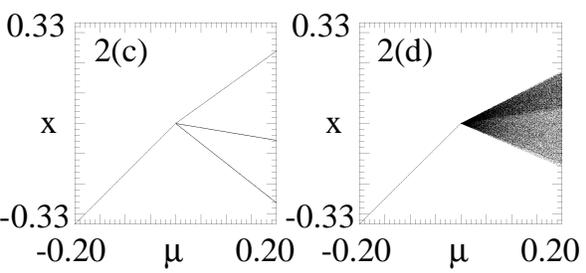
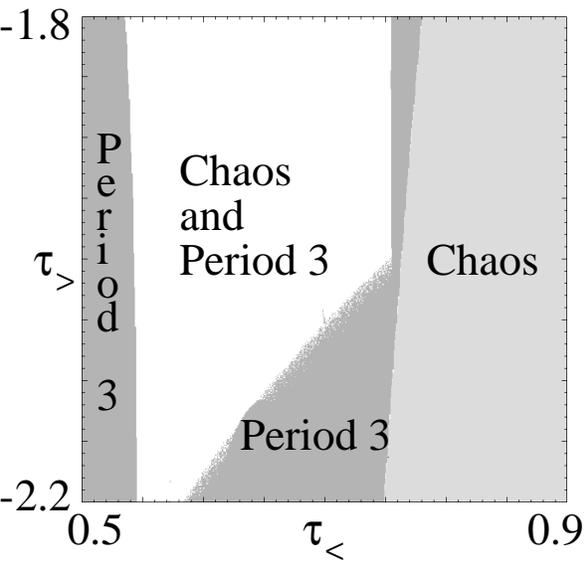
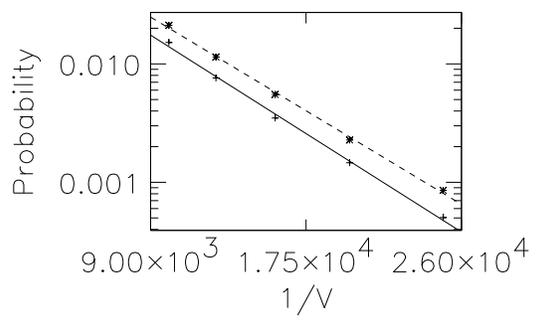
\end{document}